# On relations between probabilities under quantum and classical measurements

Andrei Y. Khrennikov [*] and Elena R. Loubenets [**] [1]

[*] *International Center in Mathematical Modelling in Physics and Cognitive Sciences, MSI, University of Vaxjo, S-35195, Sweden*
e-mail: Andrei.Khrennikov@msi.vxu.se

[**] *Moscow State Institute of Electronics and Mathematics, Technical University, Trekhsvyatitelskii Per. 3/12, Moscow 109028, Russia*
e-mail: erl@erl.msk.ru   elena@imf.au.dk

**Abstract**

We show that the so-called quantum probabilistic rule, usually introduced in the physical literature as an argument of the essential distinction between the probability relations under quantum and classical measurements, is not, as it is commonly accepted, in contrast to the rule for the addition of probabilities of mutually exclusive events. The latter is valid under all experimental situations upon classical and quantum systems.
We discuss also the quantum measurement situation that is similar to the classical one, described by Bayes' formula for conditional probabilities. We show the compatibility of the description of this quantum measurement situation in the frame of purely classical and experimentally justified straightforward frequency arguments and in the frame of the quantum stochastic approach to the description of generalized quantum measurements. In view of derived results, we argue that even under experiments upon classical systems classical Bayes' formula describes particular experimental situations that are specific for context-independent measurements. The similarity of the forms of the relation between the transformation of probabilities, which we derive in the frame of the quantum stochastic approach and in the frame of the approach, based on straightforward frequency arguments, underlines once more that projective (von Neumann) measurements represent only a special type of measurement situations in quantum physics.

## 1. Introduction

In the physical literature on quantum physics there exists a commonly accepted opinion about the essential distinction between the probabilistic rules in the classical and the quantum measurement theories. In the light of this opinion, the so called quantum probabilistic rule (see, for example, [Heisenberg (1930), Dirac (1995), Feynman (1965)])
(1) $$P_3 = P_1 + P_2 + 2\cos\theta \sqrt{P_1 P_2}$$
is put in contrast to the rule of the addition of probabilities
(2) $$P_3 = P_1 + P_2,$$

---

representing mutually exclusive events.

Another confirmation of the distinction between "classical" and "quantum" probabilities is usually found in the violation of Bayes' formulae for conditional probabilities, under the description of consecutive quantum measurements.

In the present paper we analyze such statements.

We point out that, since *we observe a quantum system by classical means, under any generalized quantum measurement the addition of probabilities of mutually exclusive events in the classical world is the main point and is included into the definition of a normalized positive operator valued (POV) measure, describing this measurement*.

We underline that the description of any quantum measurement incorporates not only the knowledge of a POV measure but also of a state of a quantum system at the instant just before a measurement. We show, in the most general settings, that the special form of the rule (1) is just connected with the fact that probabilities $P_1, P_2$ in (1) correspond to measurements, which are described by the same POV measure but carried out upon a quantum system, represented initially by different quantum states. That is why, the terms $P_1, P_2$ in (1) do not represent, under the corresponding measurement situation, the probabilities of mutually exclusive events in the classical world as well as they do not, in general, represent the probabilistic alternatives of situations in the quantum world.

We present a collection of initial states of a quantum system, where, for any type of a generalized quantum measurement, the terms $P_1, P_2$ do really represent the probabilistic alternatives of situations in the quantum world and in this case the rule (2) is valid.

We analyze a quantum measurement situation, which is similar to the classical experimental situation, described usually by Bayes' formula for conditional probabilities.

We consider the description of this quantum measurement situation in the frame of two different approaches, namely:

(a) in the frame of the approach based on the straightforward frequency approach [Khrennikov (1999-2001)]. This approach is valid for the description of both,, classical and quantum, measurement situations. The applicability of this approach in the quantum case is justified by the fact that, for processing of experimental data, obtained under an experiment upon a quantum system, the notions of classical statistics are used;

(b) in the frame of the quantum stochastic approach [Loubenets (2001); Barndorff-Nielsen and Loubenets (2002)], which can be considered as the generalization of the von Neumann approach [von Neumann (1932)] to the case of all possible quantum measurement situations.

For the considered quantum measurement situation, we point out, in particular, the quantum analog (see also in [Loubenets (2001-2)]) of classical Bayes' formula.

The compatibility of results, derived in the frames of both approaches, shows, in particular, that even under the description of consecutive measurements upon classical systems the use of classical Bayes' formula for the total probability is justified only for the special case of context independent measurements. Hence, the situation with the applicability of Bayes' rule under quantum measurements has no any connection with the argument "classical versus quantum".

## 2. On probabilistic alternatives under a quantum measurement

Let $H$ be a separable complex Hilbert space of a quantum system $S$.

Consider a generalized measurement upon $S$ with outcomes $\omega$ in the classical world of the most general possible nature, represented as points in a measurable space $(\Omega, F)$. With respect to a measurement, the space $(\Omega, F)$ is called a space of outcomes.

Let this quantum measurement be described by a normalized measure $M(\cdot)$ on $(\Omega, F)$ with values $M(B), \forall B \in F$, $M(\Omega) = I$, that are positive bounded linear operators on $H$. In the quantum measurement theory the measure $M(\cdot)$ is called a POV measure[2].

Given a state $\rho$ of a quantum system at the moment just before a measurement, the probability of the event that the outcome $\omega$ in the classical world belongs to a subset $E \in F$, is defined by the relation

(3) $$\mu_M(E; \rho) = tr\{\rho M(E)\}.$$

Notice that if $\rho$ is a pure density operator, that is, $\rho = |\varphi\rangle\langle\varphi|$, then

(4) $$\mu_M(E; |\varphi\rangle\langle\varphi|) = \|\sqrt{M(E)}\varphi\|_H^2.$$

Consider any disjoint subsets $E_1, E_2$ of $\Omega$: $E_1 \cap E_2 = \varnothing$. In the classical world, the event that $\omega \in E_1$ and the event that $\omega \in E_2$ are mutually exclusive, that is, cannot occur simultaneously under a single trial of this measurement.

By definition of a POV measure, for any disjoint subsets $E_1, E_2$,

(5) $$M(E_1 \cup E_2) = M(E_1) + M(E_2),$$

and from (3), (5) it follows that

(6) $$\mu_M(E_1 \cup E_2; \rho) = \mu_M(E_1; \rho) + \mu_M(E_2; \rho).$$

*This relation is wholly identical to the classical rule (2) of the addition of probabilities of mutually exclusive events.*

Consider now quantum measurement situations, corresponding to the derivation of the quantum probabilistic rule (1).

Specifically, let analyze the relation between the probability distributions

(7) $$\mu_M(E; \rho_i), \quad i = 1,2,3,$$

under measurements upon a quantum system $S$, which are all described by the same POV measure $M(\cdot)$, but where initial states $\rho_i, i = 1,2,3$, of $S$ are different.

Take

(8) $$\rho_i = |\varphi_i\rangle\langle\varphi_i|, \quad \varphi_i \in H, i = 1,2,3,$$

with

(9) $$\langle\varphi_i, \varphi_j\rangle = \delta_{ij}, \quad i,j = 1,2; \quad \varphi_3 = \alpha\varphi_1 + \beta\varphi_2, \quad \alpha, \beta \in C, \ |\alpha|^2 + |\beta|^2 = 1.$$

It is easy to show that, in this case, for any subset $E \in F$,

(10) $$\mu_M(E; \rho_3) = |\alpha|^2 \mu_M(E; \rho_1) + |\beta|^2 \mu_M(E; \rho_2) + 2\text{Re}\{\alpha^*\beta \langle\varphi_1, M(E)\varphi_2\rangle\}.$$

Since, due to (4), we have the following bound

$$|\text{Re}\{\alpha^*\beta \langle\varphi_1, M(E)\varphi_2\rangle\}| \leq |\alpha\beta| \sqrt{\mu_M(E; \rho_1)\mu_M(E; \rho_2)},$$

the relation (10) can be represented in the form of the quantum probabilistic rule (1):

---

[2] On the main notions of the quantum measurement theory, see [Davies (1976); Holevo (1980, 2001); Kraus (1983); Ozawa (1985); Busch, Grabowski and Lahti (1995)] and also the review sections in [Loubenets (2001-1); Barndorff-Nielsen and Loubenets (2002)].

(11) $\quad \mu_M(E;\rho_3) = |\alpha|^2 \mu_M(E;\rho_1) + |\beta|^2 \mu_M(E;\rho_2) + 2\cos\theta |\alpha\beta| \sqrt{\mu_M(E;\rho_1)\mu_M(E;\rho_2)}$,

for $\forall E \in F$.

The third term in (11) disappears if vectors $\varphi_1, \varphi_2$ satisfy the relation

$$<\varphi_1, M(\cdot)\varphi_2> = 0.$$

This implies that these vectors are mutually orthogonal with respect to the POV measure $M(\cdot)$.
The latter situation is valid, for example, for a measurement of the filter type with the set of outcomes $\Omega_{filter} = \{"\lambda^{(i)}": i=1,2.3\}$, described by the following propositions:

- "$\lambda^{(1)}$" – the quantum system is in the pure state $\rho_1$,
- "$\lambda^{(2)}$" – the quantum system is in the pure state $\rho_2$;
- "$\lambda^{(3)}$" – the quantum system is not in $\rho_1$ and not in $\rho_2$;

and elements of the POV measure, given by

$$M^{(1)} = |\varphi_1><\varphi_1|, \quad M^{(2)} = |\varphi_2><\varphi_2|,$$
$$M^{(3)} = I - |\varphi_1><\varphi_1| - |\varphi_2><\varphi_2|.$$

In general, however, the third (interference) term in (11) is present.
However, the opinion that this fact is in contradiction to the rule (2) has no any basis - since, under a measurement upon a quantum system, being initially in the state $\rho_3$, given by (8), (9), the terms $|\alpha|^2 \mu_M(E;\rho_1)$ and $|\beta|^2 \mu_M(E;\rho_2)$ *do not represent probabilities of mutually exclusive events in the classical world.*
The terms $|\alpha|^2 \mu_M(E;\rho_1)$ and $|\beta|^2 \mu_M(E;\rho_2)$ in (11) *do not also, in general, represent probabilistic alternatives in the quantum world.*
The situation is, however, quite different in the case where a generalized quantum measurement, described by a POV measure $M(\cdot)$, is carried out upon a quantum system *S*, represented initially by a density operator

$$\tilde{\rho}_3 = |\alpha|^2 \tilde{\rho}_1 + |\beta|^2 \tilde{\rho}_2, \quad |\alpha|^2 + |\beta|^2 = 1,$$

being a mixture of some density operators $\tilde{\rho}_1, \tilde{\rho}_2$.
In this case, for $\forall E \in F$,

(12) $\quad \mu_M(E;\tilde{\rho}_3) = |\alpha|^2 \mu_M(E;\tilde{\rho}_1) + |\beta|^2 \mu_M(E;\tilde{\rho}_2)$,

and the terms $|\alpha|^2 \mu_M(E;\rho_1)$ and $|\beta|^2 \mu_M(E;\rho_2)$ do really represent the probabilistic alternatives of situations in the quantum world..

## 3. On the relation between conditional probabilities

Consider a quantum measurement situation which is similar to that under classical measurements, described by Bayes' rule for conditional probabilities.
We proceed to analyze this measurement situation on the basis of two different approaches. Specifically:

- in Section 3.1, we use the quantum stochastic approach, formulated, for the description of general quantum measurements, in [Loubenets (2001); Barndorff-Nielsen and Loubenets (2002)];

- in Section 3.2, we use the approach, developed in [Khrennikov (1999-2001)], based, for measurements with different contexts, on purely classical straightforward arguments.

Notice that the latter approach is valid for both - classical and quantum measurement situations.

## 3.1 Quantum probability relation for conditional probabilities

For simplicity, we consider non-destructive quantum measurements "$A$" and "$B$", each having only two outcomes:
$$\Omega_A = \{a_i, i = 1,2\}, \quad \Omega_B = \{b_j, j = 1,2\},$$
respectively, and represented by only one measurement channel [Loubenets (2001)]. In this case, to quantum measurement "$A$" and "$B$", there exist defined uniquely (up to phase equivalence) families of bounded linear operators on $H$:

(13)          "$A$":      $\{V(a_i), \ i = 1,2, \ \sum_i V^+(a_i)V(a_i) = I_H\}$,

(14)          "$B$":      $\{W(b_j), \ j = 1,2;, \ \sum_j W^+(b_j)W(b_j) = I_H\}$,

which give the *complete statistical description* of the corresponding measurements upon a quantum system *S*.

The projective (equivalently, von Neumann) measurements corresponds to the special case where operators in (13) (or (14)) are mutually orthogonal projections, summing up to identity.

Recall [Loubenets (2001)] that, under the complete statistical description of a quantum measurement, we understand not only the specification of the probability distribution of outcomes of this measurement but also the specification of conditional posterior quantum states of *S*.

The elements $N_i^{(A)}[\cdot]$ and $N_j^{(B)}[\cdot]$, $i, j = 1,2$ of the quantum instruments[3], describing measurements "$A$" and "$B$", are given by

(15)          $N_i^{(A)}[\cdot] = V^+(a_i)[\cdot]V(a_i), \quad N_j^{(B)}[\cdot] = W^+(b_i)[\cdot]W(b_i),$

for $\forall i, j = 1,2$, and the elements of the corresponding POV measures are

(16)      $M_i^{(A)} = N_i^{(A)}[I_H] = V^+(a_i)V(a_i), \quad M_j^{(B)} = N_j^{(B)}[I_H] = W^+(b_j)W(b_j).$

Given an initial state $\rho$ of a quantum system *S*, the families

(17a)          $\{\sigma_i^{(A)}(\rho) = V(a_i)\rho V^+(a_i), \ i = 1, 2\},$

(17b)          $\{\sigma_j^{(B)}(\rho) = W(b_j)\rho W^+(b_j), \ j = 1, 2\}$ .

represent unnormalized conditional posterior states of *S*, under the measurement "$A$" and the measurement "$B$", respectively.

Due to (16) and (17), under the measurements "$A$" and "$B$", the probabilities of corresponding outcomes and the corresponding normalized conditional posterior states are given by

(18a)        $\mu^{(A)}(a_i; \rho) = tr[\rho M_i^{(A)}] = tr[\sigma_i^{(A)}(\rho)], \quad \tilde{\sigma}_i^{(A)}(\rho) = \dfrac{\sigma_i^{(A)}(\rho)}{\mu^{(A)}(a_i; \rho)},$

(18b)        $\mu^{(B)}(b_j; \rho) = tr[\rho M_j^{(B)}] = tr[\sigma_j^{(B)}(\rho)], \quad \tilde{\sigma}_j^{(B)}(\rho) = \dfrac{\sigma_j^{(B)}(\rho)}{\mu^{(B)}(b_j; \rho)}.$

We now proceed to consider different measurement situations upon a quantum system *S* represented initially by a state $\rho_0$.

The first measurement situation concerns a consecutive measurement - first "$A$" and then "$B$".

---

[3] For the notion of an instrument, cf., for example, [Holevo (2001)] and also the review sections in [Loubenets (2001-1); Barndorff-Nielsen and Loubenets (2002)],

According to the notations, introduced in (13) - (18), under this consecutive measurement, the probability $P\{(a_i,b_j)\}$ of the outcomes $a_i$ and then $b_j$ to be observed is equal to

(19) $\quad P\{(a_i,b_j)\} = \mu^{(B)}(b_j;\widetilde{\sigma}_i^{(A)}(\rho_0))\mu^{(A)}(a_i;\rho_0) = tr[W(b_j)V(a_i)\rho_0 V^+(a_i)W^+(b_j)]$,

for any $i,j = 1, 2$.

Thus, the POV measure of this consecutive measurement is represented on $\Omega_A \times \Omega_B$ by the elements

(20) $\quad M_{ij}^{(AB)} = V^+(a_i)W^+(b_j)W(b_j)V(a_i), \quad \forall i,j = 1, 2$.

If, under this consecutive measurement, the outcomes of the first measurement "$A$" are ignored, then the probability $P(b_j)$ of the outcome $b_j$ is given by

(21) $\quad P(b_j) = \sum_{i=1,2} P\{(a_i,b_j)\}$.

Introduce the density operator

(22) $\quad \widetilde{\sigma}^{(A)}(\rho_0) ==\sum_{i=1,2}\widetilde{\sigma}_i^{(A)}(\rho_0)\mu^{(A)}(a_i;\rho_0) = \sum_{i=1,2} V(a_i)\rho_0 V^+(a_i)$,

which is the unconditional posterior state of the quantum system $S$ immediately after the "$A$" measurement in case where outcomes of the "$A$" measurement are ignored.

From (19) - (22) we have:

(23) $\quad \mu^{(B)}(b_j;\widetilde{\sigma}^{(A)}(\rho_0)) := P(b_j) = \sum_{i=1,2}\mu^{(B)}(b_j;\widetilde{\sigma}_i^{(A)}(\rho_0))\mu^{(A)}(a_i;\rho_0)$.

We further refer to (23) as the *quantum analog of Bayes' formula* (see also in [Loubenets (2001-2)]).

Consider now the second measurement situation where we have to describe only measurement "$B$" upon the quantum system being initially in the state $\rho_0$.

In this case, due to (18b), the probability of an outcome $b_j$, $\forall j = 1, 2$, is given by

(24) $\quad \mu^{(B)}(b_j;\rho_0) = tr[\rho_0 W^+(b_j)W(b_j)]$.

Let now derive the *formal relation* between the probability $\mu^{(B)}(b_j;\rho_0)$, given by (24), and probabilities $P\{(a_i,b_j)\}$, $i=1,2$, introduced by (19).

Using the normality relation (see in (13)) for operators $V(a_i)$, $i=1,2$, we can rewrite (24) in the form

(25) $\quad \begin{aligned}\mu^{(B)}(b_j;\rho_0) &= \sum_{i=1,2} P\{(a_i,b_j)\} + \\ &+ \sum_{i=1,2} tr[\rho_0\{W(b_j)V^+(a_i)V(a_i)W(b_j) - V^+(a_i)W^+(b_j)W(b_j)V(a_i)\}].\end{aligned}$

Normalizing the second sum term in (25), we have

(26) $\quad \mu^{(B)}(b_j;\rho_0) = \sum_{i=1,2} P\{(a_i,b_j)\} + 2\lambda_j \sqrt{P\{(a_1,b_j)\}P\{(a_2,b_j)\}}$

with the notation

(27) $$\lambda_j = \frac{\sum_{i=1,2} tr[\rho_0\{W^+(b_j)V^+(a_i)V(a_i)W(b_j) - V^+(a_i)W^+(b_j)W(b_j)V(a_i)\}]}{2\sqrt{P\{(a_1,b_j)\}P\{(a_2,b_j)\}}}.$$

Introduce the parameters

(28) $$\gamma_{ij} = \frac{tr[\rho_0 W^+(b_j)V^+(a_i)V(a_i)W(b_j)]}{tr[\rho_0 V^+(a_i)W^+(b_j)W(b_j)V(a_i)]} = \frac{P\{(b_j,a_i)\}}{P\{(a_i,b_j)\}}, \quad \forall i,j = 1,2.$$

Here each parameter is equal to the ratio of the probabilities of the same outcomes $a_i, b_j$ but under different consecutive measurements, namely, "*B* then *A*" or "*A* then *B*".
Then

(29) $$\lambda_j = \frac{1}{2}\left[\sqrt{\frac{P\{(a_1,b_j)\}}{P\{(a_2,b_j)\}}}(\gamma_{1j} - 1) + \sqrt{\frac{P\{(a_2,b_j)\}}{P\{(a_1,b_j)\}}}(\gamma_{2j} - 1)\right], \quad \forall j = 1,2.$$

The only restriction for the parameter $\lambda_j$ is given by:

(30) $$-\frac{P\{(a_1,b_j)\} + P\{(a_2,b_j)\}}{2\sqrt{P\{(a_1,b_j)\}P\{(a_2,b_j)\}}} \leq \lambda_j \leq \frac{1 - P\{(a_1,b_j)\} - P\{(a_2,b_j)\}}{2\sqrt{P\{(a_1,b_j)\}P\{(a_2,b_j)\}}}.$$

If, for example, $P\{(a_1,b_j)\} = \frac{1}{4}$ and $P\{(a_2,b_j)\} = \frac{1}{25}$ then (30) gives $\lambda_j \in [-1,45, 3,55]$ and, consequently, in general, the absolute value of $\lambda_j$ may be more than one.

Let show, however, that in *the special case of projective (von Neumann) measurements there is only one possibility, namely,* $|\lambda_j| \leq 1$. Specifically, under a von Neumann measurement, we have

(31a) $\quad W(b_j) = |\psi_j\rangle\langle\psi_j| \quad \langle\psi_k, \psi_j\rangle = \delta_{kj}, \quad \forall k,j = 1,2,$

(31b) $\quad V(a_i) = |\varphi_i\rangle\langle\varphi_i|, \quad \langle\varphi_m, \varphi_i\rangle = \delta_{mi}, \quad \forall m,i = 1,2.$

Substituting (31) into (27) and taking also into account the representation

$$\psi_j = \sum_{k=1,2} \langle\varphi_k, \psi_j\rangle \varphi_k,$$

we derive, that, for this special type of a quantum measurement,

(32) $$|\lambda_j| = \left|\frac{\text{Re}\{\langle\sqrt{\rho_0}\varphi_1, \sqrt{\rho_0}\varphi_2\rangle \langle\varphi_2, \psi_j\rangle \langle\psi_j, \varphi_1\rangle\}}{\langle\psi_j,\varphi_1\rangle \langle\psi_j,\varphi_2\rangle \|\sqrt{\rho_0}\varphi_1\|_H \|\sqrt{\rho_0}\varphi_2\|_H}\right| \leq 1.$$

Hence, in case of projective measurements, we can use the notation
$$\lambda_j = \cos\vartheta_j.$$

From (27) it also follows that

(33) $\quad \lambda_j = 0 \quad \text{iff} \quad [W(b_j), V(a_i)] = 0, \quad \forall i,j = 1,2,$

and, in this case, the relation (26) has the form

(34) $$\mu^{(B)}(b_j;\rho_0) = \sum_{i=1,2} \mu^{(B)}(b_j; \tilde{\sigma}_i^{(A)}(\rho_0))\mu^{(A)}(a_i;\rho_0),$$

and represents the quantum analog (23) of Bayes's formula.

However, we underline once more that, in general, $|\lambda_j|$ may be more than one and, hence, may be represented as $|\lambda_j| = \cosh\theta$. We discuss this fact in detail in section 3.2.

## 3.2. *Classical (frequency) probabilistic derivation of the quantum probability relation*

In this section, we consider the relation between probabilities under the measurement situations, discussed in section 3.1, in the frame of purely classical probabilistic arguments, developed in [Khrennikov (1999-2001)]. This approach is based on the analysis of the transformation relations, induced by transitions from one complex of physical conditions, *context*, to other complexes of physical conditions, used for preparations of ensembles of physical systems.

The notion of context could be identified with the well-known notion of a *preparation procedure* (cf., for example, [Peres (1995); Busch Grabowski and Lahti (1995)]).

By using purely classical probabilistic arguments, based on calculations of relative frequencies, we classify all possible transformation relations of probabilities that could be obtained due to transition from one context to other contexts. These transitions are performed by filtration procedures. In principle, these filtration procedures, while producing new contexts, can be considered as measurements over a statistical ensemble, prepared by the old context. Purely mathematical calculations demonstrate that, besides the classical Bayes' transformation relation (described by the formula of the total probability in conventional probability theory) and the quantum rule (1) there exists a new kind of probabilistic transformation, namely the hyperbolic one.

In the case of quantum measurements, the latter transformation cannot be obtained in the frame of the von Neumann measurement postulates [von Neumann (1932)] but, however, as it was pointed out in section 3.1, it is inherent to the description of quantum measurements in the frame of the quantum stochastic approach.

Let consider a statistical ensemble $S$ of physical systems (macro or micro), produced by some reparation procedure $\mathbf{E}$. The total number of systems in $S$ is equal to N.
Suppose that there are two dichotomic physical observables, that is
(35) $$A = a_1, a_2 \quad \text{and} \quad B = b_1, b_2,$$
and that there are $n_i^a$, $i = 1, 2,$ systems in the ensemble $S$ (the sequence of trials), for which $A = a_i$, and $n_j^b$, $j = 1, 2,$ systems in $S$, for which that $B = b_j$.

Suppose also that, among those systems, for which $A = a_i$, there are $n_{ij}$, $i, j = 1, 2,$ systems for which $B = b_j$, hence,
(36) $$n_i^a = n_{i1} + n_{i2}, \quad n_j^b = n_{1j} + n_{2j}, \quad i, j = 1, 2.$$

Denote also by $S_j(A)$ and $S_j(B)$, $j=1, 2$, the sub-ensembles of $S$ with $A = a_i$ and $B = b_j$, respectively, and let
(37) $$S_{ij}(B, A) = S_i(A) \cap S_j(B)).$$
Then $n_{ij}$ is the number of elements in the ensemble $S_{ij}(B, A)$.
We would like to note that the "existence" of the property
$$\{A = a_i \text{ and } B = b_j\}$$
does not need to imply the possibility to measure this property. For example, such a measurement is impossible in case of non-commuting quantum observables.
Hence, in general, $\{A = a_i \text{ and } B = b_j\}$ is a kind of a hidden property. The $\{A = a_i \text{ and } B = b_j\}$ frequencies will be used for the further probabilistic considerations, but, of course, they will disappear from final results (this could be experimentally verified).

Physical experience says that the following frequency probabilities are well defined for any physical observables $A, B$ :

(38a) $$p_i := p_S\{A = a_i\} = \lim_{N\to\infty} p_i^{(N)}, \quad p_i^{(N)} = \frac{n_i^a}{N};$$

(38b) $$q_i := q_S\{B = b_i\} = \lim_{N\to\infty} q_i^{(N)}, \quad q_i^{(N)} = \frac{n_i^b}{N}.$$

Consider now statistical ensembles $T_i$, $i=1,2$, of physical systems, produced by some preparation procedures $\mathbf{E}_i$. In the present paper we suppose that physical systems, produced by the preparation procedure $\mathbf{E}$, pass through the additional filters $F_i$, $i=1, 2$.

In reality, this representation may induce the illusion that ensembles $T_i$, $i=1,2$ should be identified with sub-ensembles $S_i(A)$ of the ensemble $S$. However, there are no physical reasons for this identification.

The additional filter $F_1$ (and $F_2$) changes the properties of physical systems. For the ensemble $S_1(A)$ (and $S_2(A)$), the probability distribution of the property $B$ may differ from the corresponding probability distribution for the ensemble $T_1$ (and $T_2$), obtained by the filtration. Different preparation procedures produce different distributions of properties.

Our assumption on *perturbation effects* is very natural from the experimentalist (instrumentalist) point of view. This is the original viewpoint of [Heisenberg (1930)] who underlined the role of perturbations under quantum measurements.

Similar ideas were formulated in [Bohr (1935)] who emphasized the role of experimental arrangements.

The perturbation assumption implies that transitions from one context (complex of physical conditions), given by the preparation procedure $\mathbf{E}$, to other contexts, given by the preparation procedures $\mathbf{E}_j$, $j=1,2$, produce statistical perturbations of properties of physical systems, see [Khrennikov (1999-2001)], for the detailed analysis of this problem.

Suppose that there are $m_{ij}$ systems in the ensemble $T_i$, $i=1,2$, for which $B=b_j$, $j=1,2$.
Physical experience says that the following frequency probabilities are well defined:

(39) $$p_{ij} := p_{T_i}\{B = b_j\} = \lim_{N\to\infty} p_{ij}^{(N)}, \quad p_{ij}^{(N)} = \frac{m_{ij}}{n_i^a};$$

Here we assume that an ensemble $T_i$ consists of $n_i^a$ systems, $i=1,2$ and that $n_i^a = n_i^a(N) \to \infty$, $N\to\infty$. In fact, the latter assumption is true if both probabilities, $p_i$, $i=1,2$, are not equal to zero.

We remark that probabilities $p_{ij}=p_{Ti}(B=b_j)$ cannot be, in general, identified with the conditional probabilities $p_S(B=b_j/A=a_i)$. As we have noticed earlier, these probabilities are related to the statistical ensembles, prepared by the different preparation procedures, namely by $\mathbf{E}_j$, $j=1,2$, , and $\mathbf{E}$.

In our classical frequency framework we have:

(40) $$q_1^{(N)} = \frac{n_1^b}{N} = \frac{n_{11}}{N} + \frac{n_{21}}{N} = \frac{m_{11}}{N} + \frac{m_{21}}{N} + \frac{n_{11} - m_{11}}{N} + \frac{n_{21} - m_{21}}{N}$$

But since, for $i=1,2$,

(41) $$\frac{m_{1i}}{N} = \frac{m_{1i}}{n_1^a}\frac{n_1^a}{N} = p_{1i}^{(N)} p_1^{(N)}, \quad \frac{m_{2i}}{N} = \frac{m_{2i}}{n_2^a}\frac{n_2^q}{N} = p_{2i}^{(N)} p_2^{(N)},$$

we derive

(42) $$q_i^{(N)} = p_1^{(N)} p_{1i}^{(N)} + p_2^{(N)} p_{2i}^{(N)} + \delta_i^{(N)}$$

with

(43a) $$\delta_i^{(N)} = \frac{1}{N}[(n_{1i} - m_{1i}) + (n_{2i} - m_{2i})], \quad i=1,2.$$

In fact, this third term in (42) depends on the statistical ensembles $S, T_1, T_2$, and

(43b) $$\delta_i^{(N)} = \delta_i^{(N)}(S, T_1, T_2).$$

We note that, for all physical experiments, $\lim_{N\to\infty} \delta_i^{(N)}$ exists. This is the consequence of the property of the statistical stabilization of relative frequencies for physical observables (in classical, as well as, in quantum physics).
This property may be a peculiarity of nature or just a property of our measurement and preparation procedures (see [Khrennikov (2001)], for the detailed discussion on this problem). However, in any case, we always observe that, under $N\to\infty$,

(44) $$q_i^{(N)} \to q_i, \; p_i^{(N)} \to p_i, \; p_{ij}^{(N)} \to p_{ij} \;.$$

Thus, there exists the following limit

(45) $$\delta_i = \lim_{N\to\infty} \delta_i^{(N)} = q_i - p_1 p_{1i} - p_2 p_{2i},$$

and this limiting perturbation coefficient does not depend on concrete ensembles $S, T_1, T_2$ but only on the preparation procedures, that is:

(46) $$\delta = \delta(\mathbf{E}, \mathbf{E}_1, \mathbf{E}_2).$$

Suppose that ensemble fluctuations produce negligibly small (with respect to $N$) changes in properties of systems. Then

(47a) $$\delta_i^{(N)} \to 0, \; N \to \infty$$

and this asymptotics implies the classical probabilistic rule for the total probability (Bayes' formula).
In particular, this rule appears under all experiments in classical physics. Hence, the preparation and measurement procedures of classical physics produce ensemble fluctuations with asymptotics (47a).
Suppose, further, that filters $F_i, i =1,2$, produce relatively large (with respect to $N$) statistical changes in properties of systems. Then

(47b) $$\lim_{1i} \delta_i^{(N)} = \delta_i \neq 0,$$

and, in this case, we obtain the probabilistic rules which differ from the classical one.

To study carefully the behaviour of fluctuations $\delta_i^{(N)}$, we represent them as:

(48a) $$\delta_i^{(N)} = 2\lambda_i^{(N)} \sqrt{p_1^{(N)} p_{1i}^{(N)} p_2^{(N)} p_{2i}^{(N)}},$$

where the parameter

(48b) $$\lambda_i^{(N)} = \frac{1}{2\sqrt{m_{1i} m_{2i}}}[(n_{1i} - m_{1i}) + (n_{2i} - m_{2i})].$$

In (48a,b) we use the fact that

(48c) $$p_1^{(N)} p_{1i}^{(N)} p_2^{(N)} p_{2i}^{(N)} = \frac{n_1^a}{N} \cdot \frac{m_{1i}}{n_1^a} \cdot \frac{n_2^a}{N} \cdot \frac{m_{2i}}{n_2^a} = \frac{m_{1i} m_{2i}}{N^2}.$$

Furthermore, we have:

(49) $$\delta_i = 2\lambda_i \sqrt{p_1 p_{1i} p_2 p_{2i}},$$

where

(50) $$\lambda_i = \lim_{N\to\infty} \lambda_i^{(N)}, \quad i=1,2.$$

Thus, we obtained (under $N \to \infty$) the *general transformation relation between probabilities* which is induced by the transition from the ensemble $S$ (produced by $\{\mathbf{E}\}$) to ensembles $T_i$, $i=1,2$, (produced by $\mathbf{E}_i$):

$$(51) \quad q_i = p_1 p_{1i} + p_2 p_{2i} + 2\lambda_i \sqrt{p_1 p_{1i} p_2 p_{2i}}.$$

In case of *measurements upon classical objects* all coefficients $\lambda_i$ are always equal to zero.

Under quantum measurements, the situation is, in general, more complicated.
For commuting quantum observables $A$ and $B$, the coefficients $\lambda_i = 0$.
For non-commuting quantum *observables,* the coefficients $|\lambda_i|$ may, in general, be either

$$(52a) \quad |\lambda_i| < 1$$

or

$$(52b) \quad |\lambda_i| \geq 1.$$

In case (52a) we may represent the coefficient $\lambda_i$ via the parameter $\theta_i$ as

$$(53a) \quad \lambda_i = \cos\theta_i$$

Substituting then $\lambda_i = \cos\theta_i$ into (51), we get the probabilistic transformation relation

$$(53b) \quad q_i = p_1 p_{1i} + p_2 p_{2i} + 2\cos\theta_i \sqrt{p_1 p_{1i} p_2 p_{2i}},$$

which is similar to (1).
We recall the results of section 2.1 (in the frame of quantum stochastic approach) where the similar relation is valid for the case of projective (von Neumann) measurements. We remark, however, that in the frame of our approach the "phase" $\theta$ has purely probabilistic meaning as the trigonometric representation of the measure of statistical perturbations due to context transitions.

**Remark**. (Double Stochasticity and Conventional Quantum Physics) In fact, the conventional formalism of quantum mechanics gives only the probabilistic transformation relations:

$$(54) \quad q_i = p_1 p_{1i} + p_2 p_{2i} + 2\cos\theta_i \sqrt{p_1 p_{1i} p_2 p_{2i}}, \quad i=1,2,$$

where the matrix of transition probabilities $P=(p_{ij})$ is *double stochastic* [Khrennikov (1999)] and this condition can be considered as a constraint between preparation procedures $\mathbf{E}_1$ and $\mathbf{E}_2$.
However, the filters $F_i$, $i=1,2$, may produce statistical changes in properties of physical systems that are stronger and then the relation (52b) is valid. In this case we can represent rewrite $\lambda_i$ as

$$|\lambda_i| = \cosh\tilde{\theta}_i,$$

where $\tilde{\theta}_i$ are some parameters, which we may call "hyperbolic phases" and we get the following probabilistic transformation relation

$$(55) \quad q_i = p_1 p_{1i} + p_2 p_{2i} \pm 2\cosh\theta_i \sqrt{p_1 p_{1i} p_2 p_{2i}}, \quad i=1,2.$$

**Remark**. In relation (54) parameters $\theta_i$ can in principle take arbitrary values in $[0, 2\pi]$. In (55) parameters $\tilde{\theta}_i$ may take values only in some special intervals which depend on probabilities.

The relation (55) between probabilities cannot be obtained in the frame of the von Neumann measurement postulates but, as we have already demonstrated in section 3.1, this hyperbolic transformation of probabilities appears in the frame of the quantum stochastic approach to quantum measurements.